\documentclass[%
 reprint,
showpacs,
 amsmath,amssymb,
 aps,
]{revtex4-1}

\newcommand{\tens}[1]{\mbox{\textbf{\textit{\textsf{#1}}}}}

\usepackage{slashed}
\usepackage{xcolor}
\usepackage{comment}

\usepackage{graphicx}
\usepackage{dcolumn}
\usepackage{bm}

\begin{document}

\preprint{APS/123-QED}

\title{Van der Waals interactions between excited atoms in generic environments}

\author{Pablo Barcellona}
\email{pablo.barcellona@physik.uni-freiburg.de}
\affiliation{Physikalisches Institut, Albert-Ludwigs-Universit\"at
Freiburg, Hermann-Herder-Str. 3, 79104 Freiburg, Germany}

\author{Roberto Passante}
\email{roberto.passante@unipa.it}
\affiliation{Dipartimento di Fisica e Chimica,
Universit\`{a} degli Studi di Palermo and CNISM, Via Archirafi 36,
I - 90123 Palermo,
Italy \\
and INFN, Laboratori Nazionali del Sud, I-95123 Catania, Italy}

\author{Lucia Rizzuto}
\email{lucia.rizzuto@unipa.it}
\affiliation{Dipartimento di Fisica e Chimica,
Universit\`{a} degli Studi di Palermo and CNISM, Via Archirafi 36,
I - 90123 Palermo,
Italy \\
and INFN, Laboratori Nazionali del Sud, I-95123 Catania, Italy}

\author{Stefan Yoshi Buhmann}
\email{stefan.buhmann@physik.uni-freiburg.de}
\affiliation{Physikalisches Institut, Albert-Ludwigs-Universit\"at
Freiburg, Hermann-Herder-Str. 3, 79104 Freiburg, Germany}
\affiliation{Freiburg Institute for Advanced Studies,
Albert-Ludwigs-Universit\"at Freiburg, Albertstr. 19, 79104 Freiburg,
Germany}

\date{\today}

\begin{abstract}

We consider the the van der Waals force involving excited atoms in
general environments, constituted by magnetodielectric bodies.
We develop a dynamical approach studying the dynamics of the atoms and
the field, mutually coupled. When only one atom is excited, our
dynamical theory suggests that for large distances the van der Waals
force acting on the ground-state atom is monotonic, while the force
acting in the excited atom is spatially oscillating. We show how this
latter force can be related to the known oscillating Casimir--Polder
force on an excited atom near a (ground-state) body. Our force also
reveals a population-induced dynamics: for times much larger that the
atomic lifetime the atoms will decay to their ground-states
leading to the van der Waals interaction between ground-state atoms.
\pacs{34.20.Cf, 31.30.jh, 42.50.Nn, 12.20.-m}
\end{abstract}

\maketitle

\section{Introduction}
Casimir and van der Waals (vdW) forces are interactions between
neutral macroscopic bodies or atoms arising from the quantum
fluctuations of both the electromagnetic field and the atomic charges
\cite{Casimir48,CasimirPolder48}. They are responsible for many
characteristic phenomena in physics, chemistry and biology: the
deviation from ideal-gas behaviour in non-polar gases \cite{kihara},
latent heat of liquids, capillary attraction, physical absorption, and
cell adhesion \cite{nir}. Dispersion interactions have even played
an important role during the early stages of planet formation
\cite{pater}, and they are also supposed to have a fundamental role in
selective long-distance biomolecular recognition \cite{Preto}. Due to
their strong distance-dependence, they become more and more important
on the ever-decreasing scales of nanotechnology, where they lead to
the unwanted stiction of small mobile components \cite{Serry}.
In a series of ground-breaking experiments, the vdW force 
between an excited barium ion and a
mirror have been measured to high precision \cite{wilson,bushev}. 
Such experiments show an oscillating dependence of the vdW force on the ion.

We will focus on the vdW force between two atoms, in excited states
$\left| n \right\rangle $ and $\left| l \right\rangle $.
The interaction in this case is different from the interaction
between two ground-state atoms due to the possible exchange of a real
photon between the atoms. In the well-understood non-retarded regime,
that is, for distances $r$ much smaller than the wavelength of atomic
electronic transitions, one finds \cite{power3,Buhmann}
\begin{equation}
 \textbf{F}_{\text{nr}}\left(r\right) =
 - \frac{\textbf{e}_r }{4\pi^2 \varepsilon _0^2r^7}
 \sum\limits_{k,p} \frac{| \textbf{d}_{nk}^A|^2|
\textbf{d}_{pl}^B|^2 }
  {E_{k}^A-E_{n}^A + E_{p}^B-E_{l}^B}
\end{equation}
where, $\textbf{e}_r =\textbf{r}/r$, $\mathbf{d}_{nm}^A$ are the matrix-elements of the dipole operator
and $E_k$ the energy relative to the state $\left| k \right\rangle $. 
 For downward
transitions, $E_{k}^A-E_{n}^A + E_{p}^B-E_{l}^B$ can be negative and the resulting
energy positive, yielding a repulsive interaction.  Hence non-equilibrium
situations can provide repulsive vdW interactions.

The interaction at
larger separations has been object of controversies. In a first group
of works, it was predicted that the magnitude of the
retarded potential oscillates as a function of interatomic distance
\cite{Gomberoff,McLone,Philpott}. In a later group of publications it
was claimed that the retarded potential is non-oscillatory and
proportional to $1/r^2$ \cite{power,power2,power3,sherk}. The conflicting
results are due to subtle differences in treating divergent energy
denominators in the photon propagators: the poles in the real axis can
be avoided using the principal value prescription or adding
infinitesimal factors in the energy denominators and this leads to
different results. Both procedures are mathematically correct, but
they yield different physical results: a spatially oscillatory
behaviour of the interaction in the first case and a monotonically
decreasing behaviour in the second. 

A group of recent works have used dynamical approaches to address the
problem. By an appropriate time-averaging procedure \cite{manuel} or
in the limit of vanishing atomic line widths \cite{berman}, a third
result, for the vdW interaction on the excited atom, was found that
oscillates in magnitude and sign. Note that an earlier approach
based on time-dependent perturbation theory yields a non-oscillatory
result for the force on the ground-state atom that is however valid only for times shorter than the lifetime
of the excited state \cite{rizzuto}. Similar considerations about
timescales hold for the diagrammatic non-equilibrium description used
in Ref.~\cite{Haakh}.

A very recent work claims that both results, the monotonic and the
oscillating, are valid, but they describe different physical processes
\cite{milonni3}: the oscillating result is related to a coherent
exchange of excitation between the atoms, while the monotonic result
is associated to a fast loss of excitation acquired  from the
initially excited atom. Another recent work finds that both forces can
simultaneously arise in a single set-up: the vdW interaction on the
excited atom oscillates, in agreement with
\cite{manuel,berman,milonni3}, but the vdW force acting on the
ground-state atom is monotonic \cite{manuel2}. This result would imply
an apparent violation of the action--reaction principle in excited
systems in free space. However, it was shown that the momentum balance
is restored when taking the photon emitted by the excited atom into
account \cite{manuel3}. This emission being asymmetric due to the
presence of the ground-state atom, the emitted photon carries some
average momentum, so that the difference between forces on the excited
vs ground-state atoms can be interpreted as a photon recoil force.
The situation is somewhat similar to the lateral Casimir--Polder
force on an atom near a nanofibre, which is also associated with
asymmetric emission \cite{Scheel15}.

In this paper, we study the van der Waals interaction involving
excited atoms by means of a dynamical approach on the basis of the
Markov approximation. We show that the damped internal atomic dynamics
uniquely determines the oscillatory or monotonic behavior of the
retarded interaction for excited atoms. In our dynamical model, the
poles in the real axis are automatically shifted to the upper or lower
part of the complex plane, and no ad hoc choice for the imaginary
shifts in the denominators is required. We will show that, when one
atom is excited, the vdW force acting on the ground-state atom is
monotonic and the vdW interaction of the excited atom is oscillating,
in agreement with the most recent results in literature
\cite{milonni3,manuel2,manuel3,safari}. Our dynamical approach is an
alternative to the time-dependent perturbation theory \cite{manuel2},
where the behaviour of the force is determined via a time average over
rapid oscillations on time scales of the order of atomic transition
frequencies. Instead, our model allows us to study the decay-induced
dynamics on larger time scales of the order of the excited-state
lifetimes. It reveals that the force is governed by
population-induced dynamics on these scales, where for times much
larger than the lifetime of the intial atomic state the vdW
force converges to that between ground-state atoms. In addition, we
are able to account for a general environment for the two atoms, via
the classical Green tensor. 

The article is organised as follows. In Sect.~\ref{Sec2}, we present
the basic formalism describing the coupled atom--field dynamics. It is
used in Sect.~\ref{Sec3} for calculating the force between two atoms
in arbitrary excited initial states. In Sect.~\ref{Sec4}, we make the
connection of our result with the Casimir--Polder force between an
excited atom and a body of arbitrary shape. Some conclusions are given
in Sect.~~\ref{Sec5}, while in the Appendices, we present some of the
more cumbersome details of our general approach and our calculation.

\section{Atom--field dynamics}
\label{Sec2}
We consider the mutually coupled evolution of two atoms and the
medium-assisted field. The field is prepared at zero temperature, and
the atoms in generic internal states. The dynamics of the atoms can be
described with time-dependent  flip operators, defined by
$\hat A_{mn} = \left| m^A \right\rangle \left\langle n^A \right|$, where
$\left| n^A \right\rangle $ is an energy eigenstate, and similarly
$\hat B_{pq}= \left| p^B \right\rangle\left\langle q^B \right|$.

In order to evaluate the force between the two atoms we must first
solve the atom-field dynamics to obtain the flip operators and the
field operators in the Heisenberg picture. The total Hamiltonian is
the sum of three terms, the atomic and the field Hamiltonian and the
interaction term in the multipolar coupling scheme within dipole
approximation:
\begin{align}\nonumber
\hat H =& \hat H_A + \hat H_F + \hat H_{AF} \\ \nonumber
\hat H_A=& \sum\limits_n E_n^A \hat A_{nn}  + \sum\limits_n E_n^B \hat B_{nn} \\ \nonumber
\hat H_F =& \sum\limits_{\lambda  = e,m} \int \mathrm{d}^3r  \int\limits_0^\infty
\mathrm{d}\omega  \hbar \omega \mathbf{\hat f}_\lambda ^\dag \left( \mathbf{r},\omega  \right)
\cdot \mathbf{\hat f}_\lambda \left( \mathbf{r},\omega  \right)\\
\hat H_{AF} =&  - \mathbf{\hat d}^A\cdot \mathbf{\hat E}\left( \mathbf{r}_A \right)
 - \mathbf{\hat d}^B\cdot \mathbf{\hat E}\left( \mathbf{r}_B \right)
 \end{align}
where  $\mathbf{\hat f}_\lambda \left( \mathbf{r},\omega  \right)$ is the
annihilation operator for the elementary electric and magnetic
excitations of the system \cite{Buhmann-Welsch}.

Since the evolution of the whole system is unitary the commutator between two electric fields
coincides with the commutator between free fields \cite{Buhmann2,knoll} 
\begin{equation}\big[ \mathbf{\hat E}\left( \mathbf{r},\omega
\right),\mathbf{\hat E}^\dag
\left( \mathbf{r}',\omega ' \right) \big] =
 \frac{\hbar \mu _0}{\pi } \textup{Im}\tens{G}
 \left( \mathbf{r},\mathbf{r'},\omega  \right)\omega ^2\delta \left(
\omega  - \omega ' \right) \end{equation}
where $\tens{G}$ is the Green's tensor of the electromagnetic field and $\mathbf{\hat E}\left( \mathbf{r},\omega
\right)$ is the Fourier component of the electric field
$\mathbf{\hat E}\left( \mathbf{r}\right)=\int_0^\infty \text{d}\omega \mathbf{\hat E}\left( \mathbf{r},\omega
\right) +\textup{h.c.} $, Heisenberg equations for the coupled atom--field
dynamics read
\begin{gather}\nonumber
\partial_t \hat A_{mn} = \mathrm{i}\omega _{mn}^A \hat A_{mn}+ \frac{\mathrm{i}}{\hbar }
 \mathbf{\hat K}_{mn}^A \cdot \mathbf{\hat E}\left(\mathbf{r}_A \right) \\  \nonumber
 \partial_t \mathbf{\hat E} \left( \mathbf{r},\omega
\right) =
- \mathrm{i}\omega \mathbf{\hat E}\left( \mathbf{r},\omega \right)\\
+ \frac{\mathrm{i}\mu _0}{\pi }\omega ^2\left[ \textup{Im}\tens{G}
\left( \mathbf{r},\mathbf{r}_A,\omega  \right) \cdot
 \mathbf{\hat d}^A + \textup{Im}\tens{G}\left( \mathbf{r},
 \mathbf{r}_B,\omega  \right) \cdot \mathbf{\hat d}^B\right]
 \label{eqn}
\end{gather}
 where $\mathbf{\hat K}_{mn}^A = \left[ \hat A_{mn},\mathbf{\hat d}^A \right] $.
 
\begin{figure}[ht]
        \centerline{\scalebox{0.69}{\includegraphics{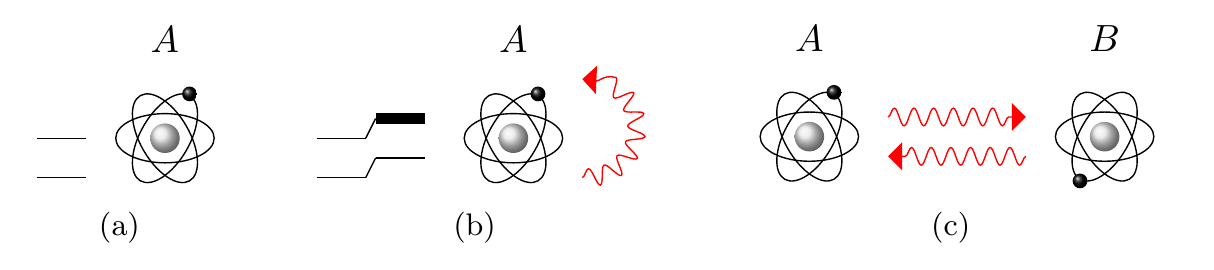}}}
        \caption{Case a) is the zero order approximation: free field and free atom. Case b) is the next order approximation: the Lamb shift of an atom
due to the emission and re-absorption of a photon. Case c) is the dispersion interaction
 between two atoms due to the exchange of two photons.}
        \label{fig1}
    \end{figure}

The electric field at the position of 
atom A consists of two terms: the
radiation reaction and the field due to the other atom B. 
As shown in the literature \cite{ackerhalt,Buhmann2}, the radiation
reaction field gives rise to frequency shifts and spontaneous decay
for atom A, see Figs.~\ref{fig1}(a) and (b). 
We thus renormalize
the field by splitting off the radiation reaction
\begin{multline}
\left\langle \partial_t \hat A_{mn} \right\rangle  = \left[ \mathrm{i}\tilde \omega _{mn}^A - \left(
 \Gamma _n^A + \Gamma _m^A \right)/2 \right]\left\langle \hat A_{mn}
  \right\rangle  \\
  + \frac{\mathrm{i}}{\hbar }\left\langle \mathbf{\hat K}_{mn}^A \cdot
   \mathbf{\hat E}_\slashed{A}\left( \mathbf{r}_A,t \right) \right\rangle
   \label{eqn21}
   \end{multline}
where $m \ne n$ and the expectation value
\mbox{$\langle\ldots\rangle$} is taken over atomic state and the field
thermal state. $\mathbf{\hat E}_\slashed{A}\left( \mathbf{r}_A,t \right) $
is the sum of the free electric field and the source field of the atom B, $\tilde \omega
_{mn}^A$ the (second-order) Lamb-shifted atomic frequencies and $\Gamma _n^A$ the
decay rates.

\section{Van der Waals interaction between two excited atoms}
\label{Sec3}

We consider two atoms $A$ and $B$ that are initially prepared in excited
energy eigenstates $\left| i_A \right\rangle$, $\left| i_B
\right\rangle$  of the free atomic Hamiltonian
($p_n^A(0)=\delta_{ni_A}, p_l^B(0)=\delta_{li_B}$). 
These initial states are not eigenstates of the total Hamiltonian and
thus the atomic states evolve in time yielding a time-dependent vdW
force (population-induced dynamics). As time progresses, the lower
lying levels $n \leqslant i_A$,$l \leqslant i_B$ will become
populated.

To find the vdW force between on, say, atom $A$, we calculate the 
Lorentz force in electric-dipole approximation acting on $A$ which
is due to the field $\mathbf{E}_\slashed{A}\left(
\mathbf{r}_A,t \right) $ emitted by the other atom B:
\begin{equation}\mathbf{F}_A (\mathbf{r}_A,\mathbf{r}_B,t) = \nabla
_A\big\langle \mathbf{\hat d}^A \cdot \mathbf{\hat
E}_\slashed{A}\left( \mathbf{r}_A,t \right) \big\rangle, 
 \end{equation}
where expectation value is taken over atomic and field states.

For weak atom--field coupling,  corresponding in an expansion of the
Hamiltonian in powers of the coupling strengths $\textbf{d}$, we can
apply the Markov approximation to find (see Appendix):
\begin{multline}
\textbf{F}_A\left( \mathbf{r}_A,\mathbf{r}_B,t \right)=
 \frac{\mu _0^2}{2\pi ^2\hbar} \sum\limits_{n \leqslant i_A } \sum\limits_{l \leqslant i_B }p_n^A(t)p_l^B(t) \sum\limits_{k,p} \\ \times \int\limits_0^\infty
  \mathrm{d}\omega \int\limits_0^\infty  \mathrm{d}\omega '   \omega ^2\omega '^2 
  \nabla _A\left\{ \mathbf{d}_{nk}^A \cdot \text{Im}\tens{G}\left( \mathbf{r}_A,\mathbf{r}_B,\omega  \right) \right. \cdot \mathbf{d}_{pl}^B\\
 \left.  \times \mathbf{d}_{lp}^B \cdot \text{Im}\tens{G}\left( \mathbf{r}_B,\mathbf{r}_A,\omega ' \right) \cdot \mathbf{d}_{kn}^A \right\}\sum\limits_{i = 1}^{16} \frac{1}{D_i} +\text{h.c.}
    \label{pa500}
\end{multline}
where $p_n^A(t) = \big \langle \hat A_{nn}\left( t \right) \big\rangle $, $p_l^B(t) = \big \langle \hat B_{ll}\left( t \right) \big\rangle$ represent the atomic populations of states $\left| n \right\rangle $ and $\left| l \right\rangle $.

 The energy denominators $D_i$ are listed in Table 1.
\begin{table}[h]
\begin{tabular}{ |l|l| }
  \hline
  \multicolumn{2}{|c|}{Energy denominators} \\
  \hline   
 & \\ 
$D_1$ & $( \omega^{(-)}  + \omega _{kn}^{A(-)} )( \omega ' + \omega _{pl}^{B(-)} )( \omega _{kn}^{A(-)} + \omega _{pl}^{B(-)}  )$ \\
$D_2$ & $(\omega^{(-)}  + \omega _{kn}^{A( - )})(\omega'  + \omega _{pl}^{B( + )})(\omega _{kn}^{A( - )} - \omega _{pl}^{B( + )})$ \\
$D_3$ & $(\omega^{(-)}   - \omega _{kn}^{A( + )})(\omega'  + \omega _{pl}^{B( - )})(\omega _{kn}^{A( + )} - \omega _{pl}^{B( - )})$ \\
$D_4$ & $( \omega^{(-)}  - \omega _{kn}^{A(+)} )( \omega ' + \omega _{pl}^{B(+)}  )(\omega _{kn}^{A(+)} + \omega _{pl}^{B(+)})$ \\
$D_5$ & $( \omega^{(-)}  + \omega _{kn}^{A(-)} )( \omega ' + \omega _{kn}^{A(-)}  )( \omega _{kn}^{A(-)} + \omega _{pl}^{B(-)} )$ \\
$D_6$ & $ - (\omega^{(-)}  - \omega _{kn}^{A( + )})(\omega ' + \omega _{kn}^{A( + )})(\omega _{kn}^{A( + )} - \omega _{pl}^{B( - )})$ \\
$D_7$ & $ - ( \omega^{(-)}  + \omega _{kn}^{A(-)}  )( \omega ' + \omega _{kn}^{A(-)}  )( \omega _{kn}^{A(-)} - \omega _{pl}^{B(+)}  )$ \\
$D_8$ & $  (\omega^{(-)}  - \omega _{kn}^{A(+)}  )( \omega ' + \omega _{kn}^{A(+)})( \omega _{kn}^{A(+)} + \omega _{pl}^{B(+)}  )$ \\
 \hline
\end{tabular}
\begin{tabular}{ |l|l| }
  \hline
  \multicolumn{2}{|c|}{} \\
  \hline   
 & \\ 
$D_9$ & $( \omega^{(-)}   + \omega ' )( \omega^{(-)}   + \omega _{kn}^{A(-)}  )( \omega ' + \omega _{pl}^{B(-)}  )\qquad  \quad$  \\
$D_{10}$ & $( \omega^{(-)}   - \omega ' )( \omega^{(-)}   + \omega _{kn}^{A(-)})( \omega ' + \omega _{pl}^{B(+)}  )$\\
$D_{11}$ & $  - (\omega^{(-)}   + \omega ')(\omega^{(-)}   - \omega _{kn}^{A( + )})(\omega ' + \omega _{pl}^{B( - )})$ \\
$D_{12}$ & $ - (\omega^{(-)}   - \omega ' )(\omega^{(-)}   - \omega _{kn}^{A(+)} )(\omega ' + \omega _{pl}^{B(+)}  )$ \\
$D_{13}$ & $( \omega^{(-)}   + \omega ' )( \omega^{(-)}   + \omega _{kn}^{A(-)}  )( \omega^{(-)}   + \omega _{pl}^{B(-)}  )$  \\
$D_{14}$ & $ - ( \omega^{(-)}   - \omega ' )( \omega^{(-)}   + \omega _{kn}^{A(-)}  )(\omega^{(-)}   + \omega _{pl}^{B(-)}  )$ \\
$D_{15}$ & $ - ( \omega^{(-)}  + \omega ' )( \omega^{(-)}   - \omega _{kn}^{A(+)}  )( \omega^{(-)}   + \omega _{pl}^{B(-)} )$  \\
$D_{16}$ & $(\omega^{(-)}   - \omega ' )( \omega^{(-)}   - \omega _{kn}^{A(+)}  )(\omega^{(-)}   + \omega _{pl}^{B(-)})$ \\
  \hline
\end{tabular}
 \caption{Energy denominators. In this table, $\omega _{kn}^A$
represents the transition frequency between the virtual state $\left|
k \right\rangle $ and the excited state $\left| n \right\rangle$,
while  $\omega _{pl}^B$ represents the  transition frequency between
the virtual state $\left| p \right\rangle $ and $\left| l
\right\rangle $.
 Furthermore $\omega _{kn}^{A\left(  \pm  \right)}
= \omega_{kn}^A \pm \text{i}\left( \Gamma _k^A + \Gamma _n^A
\right)/2$, $\omega _{pl}^{B\left(  \pm  \right)}
= \omega _{pl}^B \pm \text{i}\left( \Gamma _p^B + \Gamma _l^B
\right)/2$ and $\omega ^{\left(  \pm  \right)} = \omega  \pm
\text{i}\epsilon$ with $\epsilon$ infinitesimal factor. $\Gamma$ is
the atomic line-width.}
\end{table}\\
Due to our dynamical
treatment of the atom--field coupling, the result
explicitly depends on atomic damping constants or line widths, and
also an infinitesimal damping for the photon frequency $\omega$.
These factors uniquely ensure the convergence of time-integrals.

For excited atoms the energy denominators can exhibit poles, for
photon frequencies being resonant to the atomic ones.
According to time-independent perturbation theory these poles
would be situated on the real-frequency axis with the mentioned
resulting ambiguities. In our dynamical approach, with the inclusion of the atomic line widths, the poles are automatically shifted to the lower or 
upper part of the complex plane leading to unique resonant contributions.

The total vdW force acting on A consists in two terms, a non-resonant contribution arising from virtual photons exchange, and a resonant contribution
which corresponds to a possible emission of real photons by the excited atoms:
\begin{equation}
 \mathbf{F}_A\left( \mathbf{r}_A,\mathbf{r}_B,t \right) = \mathbf{F}_A^{\mathrm{nr}}\left( \mathbf{r}_A,\mathbf{r}_B,t \right) + \mathbf{F}_A^{\mathrm{r}}\left( \mathbf{r}_A,\mathbf{r}_B,t \right) 
\end{equation}
In the limit of vanishing line-widths, the non-resonant contribution in an arbitrary magnetoelectric environment reads (see Appendix):
\begin{multline}
\mathbf{F}_A^{\mathrm{nr}}\left( \mathbf{r}_A,\mathbf{r}_B,t \right) =
\frac{\hbar \mu _0^2}{2\pi }\int\limits_0^\infty  \text{d} \xi \xi ^4
\nabla _A \text{Tr}\left\{ \bm{\alpha} _A\left( \text{i}\xi  \right)
\right.\\ 
  \cdot \tens{G}\left( \mathbf{r}_A,\mathbf{r}_B,\text{i}\xi
\right)\left.  \cdot \bm{\alpha} _B\left( \text{i}\xi  \right) \cdot
\tens{G}\left( \mathbf{r}_B,\mathbf{r}_A,\text{i}\xi  \right) \right\}
  \label{nonresonant}
\end{multline}
where we have defined the following polarizabilities of the initially excited atoms:
\begin{align}\nonumber
\bm{\alpha} _A\left( \omega  \right) &=\frac{1}{\hbar} \sum\limits_{n
\leqslant i_A} p_n^A (t)\sum\limits_k \left(
\frac{\mathbf{d}^A_{kn}\mathbf{d}^A_{nk}}{\omega _{kn}^A + \omega } +
\frac{\mathbf{d}_{nk}^A\mathbf{d}^A_{kn}}{\omega _{kn}^A - \omega }
\right)\\
\bm{\alpha} _B\left( \omega  \right) &=\frac{1}{\hbar} \sum\limits_{l
\leqslant i_B} p_l^B (t)\sum\limits_p \left(
\frac{\mathbf{d}^B_{pl}\mathbf{d}^B_{lp}}{\omega _{pl}^B + \omega } +
\frac{\mathbf{d}_{lp}^B\mathbf{d}^B_{pl}}{\omega _{pl}^B - \omega }
\right)
\end{align}
The resonant contribution reads:
\begin{multline}
  \mathbf{F}_A^{\mathrm{r}}\left( \mathbf{r}_A,\mathbf{r}_B ,t\right)=\\
    \mu _0^2 \sum\limits_{n \leqslant i_A}  p_n^A(t)\sum\limits_{k < n} \nabla _A \operatorname{Re} \left\{ \left( \omega _{nk}^A \right)^4 \right.\\ 
\times \mathbf{d}_{nk}^A \cdot \tens{G}\left(
\mathbf{r}_A,\mathbf{r}_B,\omega _{nk}^A \right) \cdot
\bm{\alpha}_B\left( \omega _{nk}^A \right)\left.  \cdot \tens{G}\left(
\mathbf{r}_B,\mathbf{r}_A,\omega _{nk}^A \right) \cdot
\mathbf{d}_{kn}^A \right\}\\
+\mu _0^2 \sum\limits_{l \leqslant i_B}  p_l^B(t)\sum\limits_{p < l}
\nabla _A \left\{ \left( \omega _{lp}^B \right)^4 \right.\\ 
\times \left.  \mathbf{d}_{lp}^B \cdot \tens{G}\left(
\mathbf{r}_B,\mathbf{r}_A,\omega _{lp}^B \right) \cdot
\bm{\alpha}_A\left( \omega _{lp}^B \right) \cdot \tens{G}^*\left(
\mathbf{r}_A,\mathbf{r}_B,\omega _{lp}^B \right) \cdot
\mathbf{d}_{pl}^B \right\}\\
\label{resonant}
\end{multline}
For large distances the resonant contribution dominates over the non-resonant one. Two terms, one oscillating and one monotonic, are involved in the resonant contribution. Their behavior can be seen explicitly for isotropic atoms in free space:
\begin{multline}\textbf{F}_A^\mathrm{r} \left( r,t \right) =
- \frac{1}{12\pi ^2\varepsilon _0^2r^7}\,\textbf{e}_r \sum\limits_{n
\leqslant i_A }p_n^A(t)
\sum\limits_{k< n} {\left| \mathbf{d}_{nk}^A \right|}^2\\
 \times \bm{\alpha} _B\left(\omega _{nk}^A \right) \left[ \left( 9 - 16x_{nk}^2 + 3x_{nk}^4 \right)\cos \left( 2x_{nk} \right)\right.\\
\left. + \left( 18x_{nk} - 8x_{nk}^3+x_{nk}^5 \right)\sin \left( 2x_{nk} \right) \right]\\
 - \frac{1}{12\pi ^2\varepsilon _0^2r^7}\, \textbf{e}_r
\sum\limits_{l \leqslant i_B} p_l^B (t)\sum\limits_{p < l} \left|
\mathbf{d}_{lp}^B \right|^2  \\
 \times \bm{\alpha} _A\left( \omega _{lp}^B \right)\left( 9 +
2y_{lp}^2 + y_{lp}^4 \right)\end{multline}
where $x_{nk} = r  \omega _{nk}^{A}/c$ and $y_{lp} = r  \omega _{lp}^{B}/c$, $\textbf{e}_r =\textbf{r}/r$.
When both atoms are excited, the monotonic and oscillating results
both contribute and can be attributed to different physical processes
\cite{milonni3}: the oscillating result is related to a reversible
exchange of excitation (``pendulation'') and the
monotonic form with an effectively irreversible (Forster) excitation transfer.

When only one atom is excited, the force acting on the excited atom is oscillating; on the other hand, the force acting on the ground-state atom is monotonic, coherently with the perturbative result in \cite{rizzuto}. 
This implies a violation of the action-reaction principle in excited systems in free space. The interaction is accompanied by the transfer of linear momentum
to the electromagnetic vacuum; this momentum is ultimately released through  directional spontaneous emission of the excited atom \cite{manuel3}.

\begin{figure}[ht]
        \centerline{\scalebox{0.41}{\includegraphics{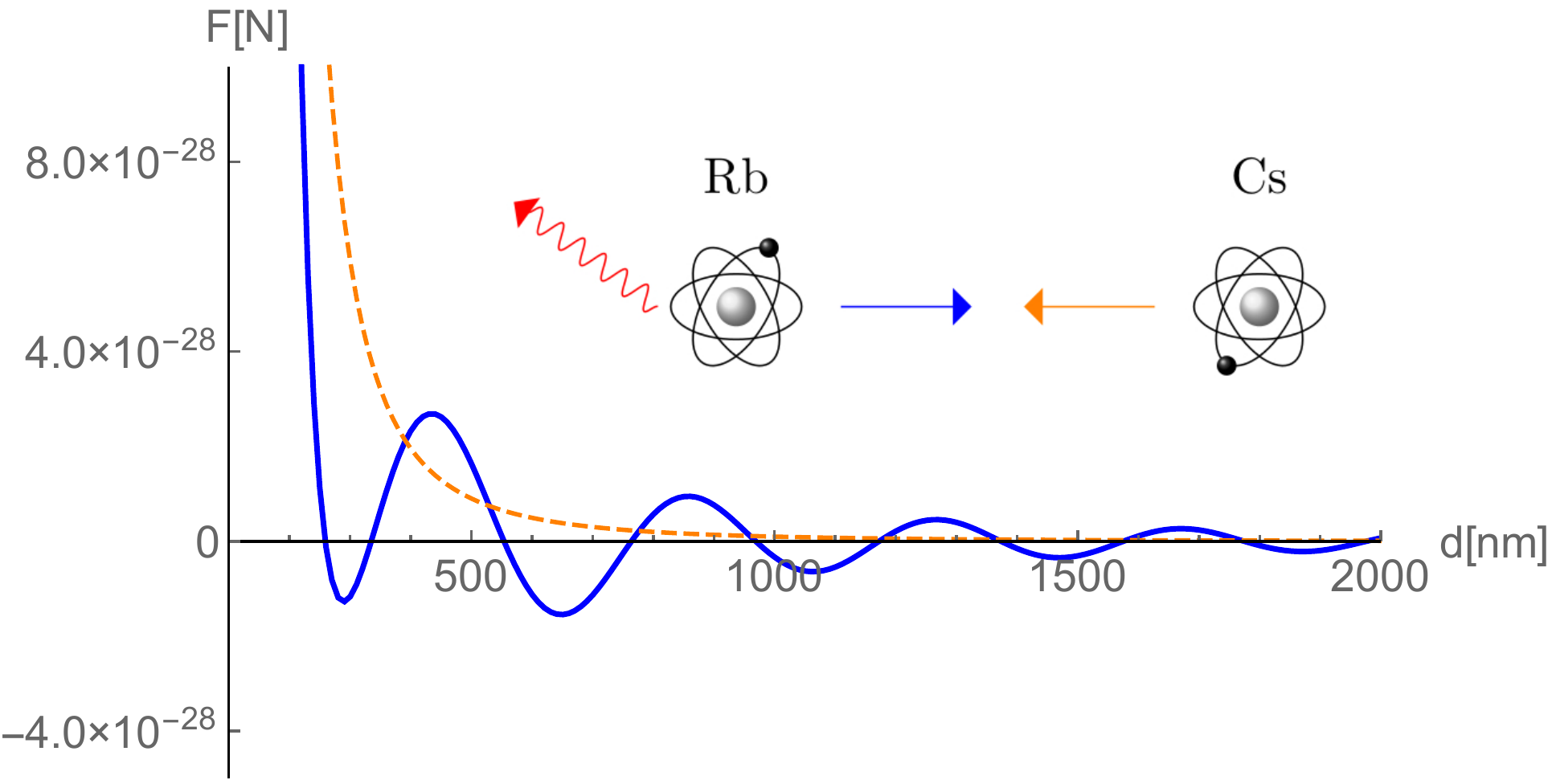}}}
        \caption{vdW interaction between one Cesium atom in the
ground state ($5^2S_{1/2}$) and an excited Rubidium atom ($5^2P_{1/2}$). The thick line represents the force on Rubidium and the dashed one that on Cesium. }
        \label{fig2}
    \end{figure}

In Fig.~(\ref{fig2}), we show the vdW force acting on a rubidium atom and on a Cesium atom in free space, the Rubidium atom
being in the excited state $5^2P_{1/2}$ and the Cesium atom in the ground-state $5^2S_{1/2}$ (see \cite{steck}); the force is represented for times
much shorter than the atomic lifetime and much larger than the inverse of the atomic frequency, so that the populations of the states may be considered constant and 
the atomic dynamical self-dressing is not present.
 At large distances the resonant term dominates and the
force on the excited atom shows Drexhage-type oscillations with an amplitude $r^{-2}$.
 The force acting on the ground-state atom is monotonic. At
small distances, we find a non-oscillating repulsive force for both atoms.
    
\begin{figure}[ht]
        \centerline{\scalebox{0.71}{\includegraphics{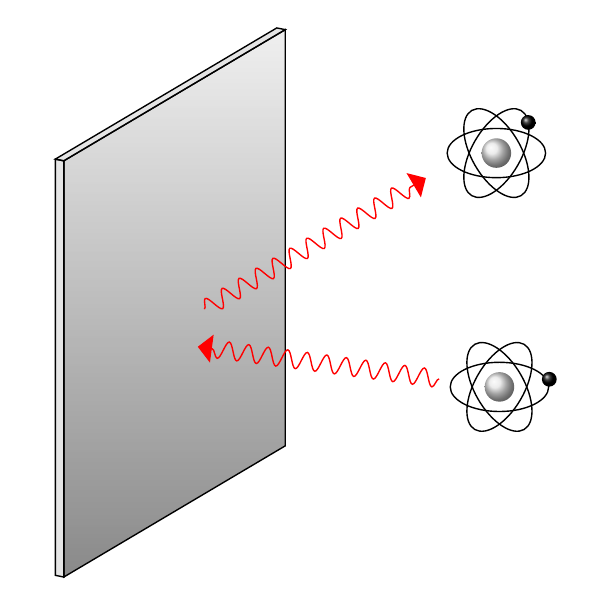}}}
        \caption{Body-assisted vdW interaction: the exchanged photons can be reflected by the body's surface.}
        \label{fig3}
    \end{figure}

    However our theory is more general because it includes the presence of general environments for the two atoms, like magnetodielectric bodies.
    Many differences arise in this more general case. Firstly the interaction can be described as a two-photon process, where the photons can be reflected by the body's surface (see Fig. \ref{fig3} ); this reflection is mathematically described in our formalism by the scattering Green tensor, which is known for many geometries and magnetodielectric properties. 
    Secondly due to the presence of the additional body the action-reaction principle is also violated for ground-state atoms, with the interaction being accompanied by the transfer of linear momentum to the body.  Lastly the total force acting on one molecule is not parallel to the interparticle separation vector.
    
\begin{figure}[ht]
        \centerline{\scalebox{0.71}{\includegraphics{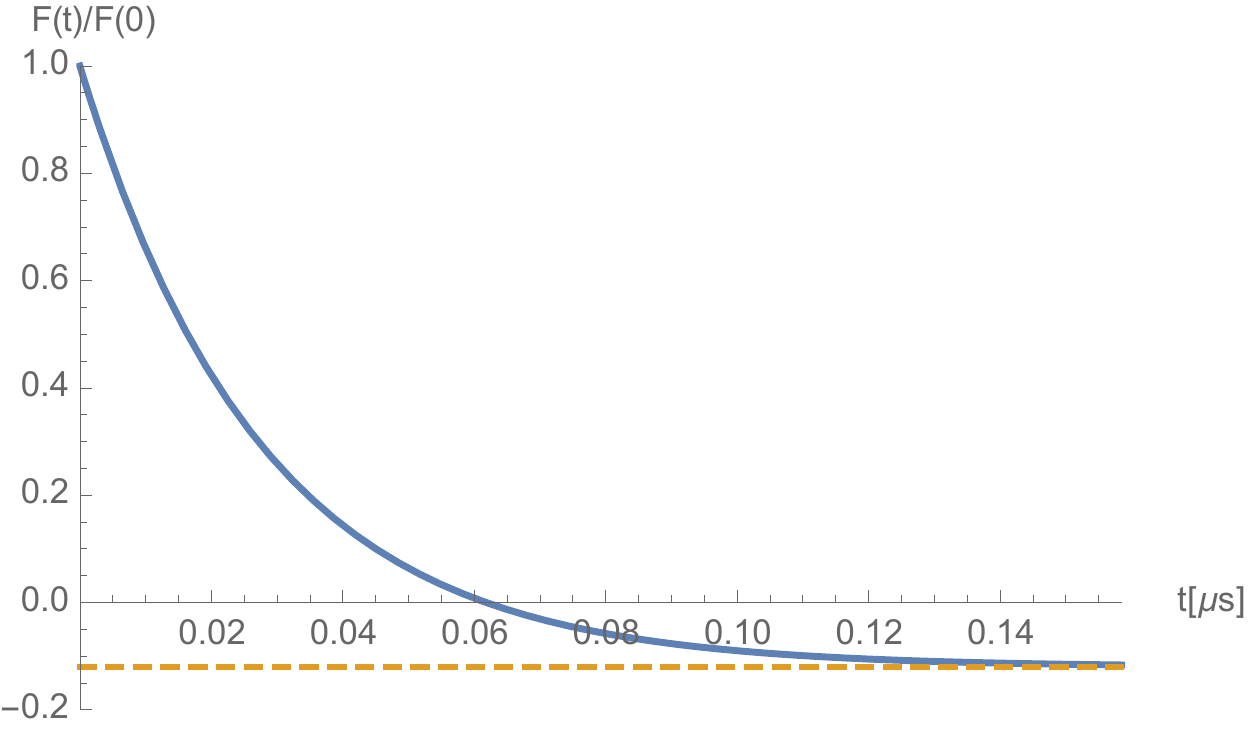}}}
        \caption{Population-induced dynamics for the vdW force acting on the excited Rubidium atom (thick line). The Rubidium atom is in the $5^2P_{1/2}$ state, while  the Cesium is in the ground-state. The distance between the atoms is $r=10nm$.}
        \label{fig4}
    \end{figure}

We see that the resonant contribution vanishes for times much larger than the atomic lifetimes ($\mathbf{F}_A^{\text{r}} \propto p_n^A\left( t \right) = e^{ - \Gamma _n^At}$), when the atoms have decayed to the ground-state. 
Fig.~(\ref{fig4}) represents this population-induced dynamics for the force acting on the excited Rubidium at a given distance.
We see that for time much larger than the atomic lifetime the force converges to the ground-state force, which is attractive. For times much smaller than the atomic lifetime the force is repulsive and roughly one order of magnitude larger than the ground-state force.

As stated above, the resonant force on the excited atom may be associated with photon recoil due to spontaneous emission. The fact
that this force is stronger at small times can be understood from its ensemble-average nature: the probability of photon emission (and hence recoil) is highest
for small times where a large fraction of the ensemble atoms are still in their excited state.

\section{Comparison to Casimir--Polder force}
\label{Sec4}    

Let us compare our result with the
experimentally observed single-atom Casimir--Polder force. If an
initially excited atom $A$ is placed near a magnetodielectric body,
the resonant contribution, associated with a possible emission of a
real photon, reads \cite{Buhmann2}:
\begin{multline}
\mathbf{F}_A^r\left( \mathbf{r}_A,t \right) = \mu _0\sum\limits_{n
\leqslant i_A} p_n^A\left( t \right) \sum\limits_{k < n} \left( \omega
_{nk}^A \right)^2	\\	\times \nabla _A \text{Re} \left\{
\mathbf{d}_{nk}^A \cdot \tens{G}^1\left(
\mathbf{r}_A,\mathbf{r}_A,\omega _{nk}^A \right) \cdot
\mathbf{d}_{kn}^A \right\}
\end{multline}
where $\tens{G}^1$ is the body's scattering Green's tensor. If the
body is made up of ground-state atoms with polarizability $\bm{\alpha}
_B\left( \omega  \right)$ and positions $\textbf{r}_B$ and number
density $\eta(\mathbf{r})$, it can be expressed in terms of a
leading-order Born expansion \cite{born}:
\begin{multline}
\tens{G}^1\left( \mathbf{r}_A,\mathbf{r}_A,\omega  \right)=
\mu _0\omega ^2\int \text{d}^3r_B\,\eta(\mathbf{r}_B) \tens{G}^0\left(
\mathbf{r}_A,\mathbf{r}_B,\omega  \right)\\  \cdot \bm{\alpha}_B\left(
\omega \right) \cdot \tens{G}^0\left( \mathbf{r}_B,\mathbf{r}_A,\omega
\right) + ...
\label{Green}
\end{multline}
where $\tens{G}^0$ is the free-space Green's tensor. The substitution
of this expansion into the single-atom Casimir--Polder force leads to
a resonant force
\begin{multline}
\label{CP}
\mathbf{F}_A^r\left( \mathbf{r}_A,t \right) =
\int \text{d}^3r_B\,\eta(\mathbf{r}_B)
\mathbf{F}_A^{\mathrm{r}}\left( \mathbf{r}_A,\mathbf{r}_B \right)\\
=\int \text{d}^3r_B\,\eta(\mathbf{r}_B)\mu _0^2 \sum\limits_{n
\leqslant i_A}  p_n^A(t)\sum\limits_{k < n} \nabla _A
\operatorname{Re} \left\{ \left( \omega _{nk}^A \right)^4 \right.
\mathbf{d}_{nk}^A \\ \cdot \tens{G}^1\left(
\mathbf{r}_A,\mathbf{r}_B,\omega _{nk}^A \right) \cdot
\bm{\alpha}_B\left( \omega _{nk}^A \right)\left.  \cdot
\tens{G}^1\left(
\mathbf{r}_B,\mathbf{r}_A,\omega _{nk}^A \right) \cdot
\mathbf{d}_{kn}^A \right\}
\end{multline}
on the excited atom which is simply the sum over the (oscillating)
resonant forces~(\ref{resonant}) on the excited atom due to
the ground-state atoms constituting the body.
Note that a monotonous force contribution is absent from the
single-atom Casimir--Polder force~(\ref{CP}), as the atoms in the body
are not excited. In the present combination of an excited atom
interacting with a ground-state atom, we would expect the force on the
body to contain a monotonous Casimir--Polder force component. However,
the force on the body is usually not considered in the context of
Casimir--Polder physics due to the strongly asymmetric mass ratio.

\section{Conclusions and outlook}
\label{Sec5}

Our dynamical theory has allowed us to study the vdW force involving
excited atoms in generic environments. It is able to give a unique
answer to the old puzzle whether the respective interaction is
oscillating or monotonic, without recourse to ad hoc assumptions or
prescriptions.

When one atom is excited we have shown that the van der Waals force
acting on the excited atom indeed shows Drexhage-type oscillations,
while the force acting on the ground-state atom is monotonic. We have
explicitly demonstrated that the oscillating force is consistent with
the respective Casimir--Polder force between an excited atom and a
ground-state body. On the contrary, the monotonic forces components
cannot be deduced from the atom--body force in this way, because they
act on the atoms inside the body whereas Casimir--Polder calculations
are usually restricted to calculating the force on the single atom in
front of the body.

The oscillating force on the excited atom could have profound
implications on the spatial correlations of excited atomic ensembles,
in particular for Rydberg systems. In addition, both the oscillating
and monotonous force components are expected to arise in waveguides as
recently studied in
Refs.~\cite{Shamoon13,Shamoon14,ScheelHaakh,Farina}.
It could be also interesting to generalize our model to include finite
temperature, by chancing the fluctuation relations of the
electromagnetic field, and to consider many-body vdW forces.

\begin{acknowledgements}
We thank M. Donaire, H.~Haakh, J. Hemmerich and P. W. Milonni for discussions.
RP and LR gratefully acknowledge financial support by the
Julian Schwinger Foundation and by MIUR. SYB and PB are
grateful for support by the DFG (grants BU 1803/3-1 and GRK
2079/1) and the Freiburg Institute for Advanced Studies.
\end{acknowledgements}

\section{Appendix}
\subsection{Perturbative expansion of the force operator}
We consider the dynamics of an operator $\hat \rho(t)$, which is a superposition of operators
$\hat O_n(t)$ with complex coefficients $f_n(t)$:
\begin{equation}
\hat \rho \left( t \right) = \sum\limits_n f_n \left( t \right) \hat O_n\left( t \right)
\end{equation}
and we introduce a time limit which acts only on the operators:
\begin{equation}
\left. \hat \rho \left( t \right) \right|_{t \to t_1} = \sum\limits_n f_n \left( t \right) \hat O_n\left( t_1 \right)
\end{equation}
The operators $\hat O_n(t)$ evolve dynamically according to the Heisenberg equations:
\begin{equation}
\partial _t \hat  O_n\left( t \right) = \frac{1}{\text{i}\hbar }\big[ \hat O_n\left( t \right),\hat H\left( t \right) \big]
\end{equation}
where $\hat H$ is the total Hamiltonian. This equation can be integrated from the initial time $t_0$
to a given time $t$:
\begin{equation}
 \hat O_n\left( t \right) = \hat O_n\left( t_0 \right) + \frac{1}{\text{i}\hbar }\int\limits_{t_0}^t \text{d}t_1\big[ \hat O_n\left( t_1 \right),\hat H\left( t_1 \right) \big]
\end{equation}
This equation shows that the dynamical evolution of the operators $\hat \rho$ is:
\begin{equation}
\hat \rho \left( t \right) = \left. \hat \rho \left( t \right) \right|_{t \to t_0} + \frac{1}{\text{i}\hbar }\int\limits_{t_0}^t \text{d}t_1\big[ \left. \hat  \rho \left( t \right) \right|_{t \to t_1},\hat H\left( t_1 \right) \big]
\label{rho}
\end{equation}
The vdW force operator acting on the atom A, due to the presence of other atoms, is:
\begin{equation}\hat{\mathbf{F}}_A (t) = \nabla \big[ \hat{\mathbf{d}}_A \cdot \hat{\mathbf{E}}_\slashed{A}\left(\mathbf{r},t \right) \big]_{\mathbf{r} = \mathbf{r}_A}
\label{multham} \end{equation}
where the field $\hat{\mathbf{E}}_\slashed{A}\left( \mathbf{r}_A,t \right)$ represents the total electric field, excluding the radiation reaction of atom A.
 Using Eq.~(\ref{rho}) we find the dynamical equation for the force:
\begin{equation}
\hat{\mathbf{F}}_A(t) =  \hat{\mathbf{F}}_A(t) \big|_{t \to t_0} + \frac{1}{\text{i}\hbar }\int\limits_{t_0}^t \text{d} t_1 \Big[  \hat{\mathbf{F}}_A(t) \big|_{t \to t_1},\hat H\left( t_1 \right) \Big]\\
\end{equation}
This equation can be reiterated considering now the dynamics of the commutator $\Big[  \hat{\mathbf{F}}_A(t) \big|_{t \to t_1},\hat H\left( t_1 \right) \Big]$, which is a superposition of operators at the time $t_1$. Therefore, for weak coupling, we can construct a perturbative expansion $\hat {\mathbf{F}}_A(t)$ in terms of operators at the initial time
$t_0$:
\begin{multline}
\hat{\mathbf{F}}_A\left( t \right) = \hat{\mathbf{F}}_A\left( t \right) \big|_{t \to t_0}\\
 + \frac{1}{\text{i}\hbar }\int\limits_{t_0}^t \text{d}t_1 \big[ \hat{\mathbf{F}}_A \left( t \right) \big|_{t \to t_1},\hat H\left( t_1 \right) \big]_{t_1 \to t_0}  \\
 + \left( \frac{1}{\text{i}\hbar } \right)^2\int\limits_{t_0}^t \text{d}t_1\int\limits_{t_0}^{t_1} \text{d}t_2 \\
\times   \Big[  \big[  \hat{\mathbf{F}}_A \left( t \right) \big|_{t \to t_1},\hat H\left( t_1 \right)  \big]_{t_1 \to t_2},\hat H\left( t_2 \right)  \Big] _{t_2 \to t_0} + ...
 \label{eqrho}
\end{multline}

In our model the electric field and the flip-operators of the two
atoms are the dynamical
variables of the system. Two different time-scales are observed for the dynamical variables; there is a fast free dynamics and a much slower dynamics due to the interaction between the atoms and the field. For example the free evolution of the flip operators is on time scales of $\omega_0^{-1}=10^{-15}s$ while the dynamics due to the interaction
is on time scales of $\Gamma^{-1}=10^{-9}s$. We define new dynamical variables according the formulas:
\begin{align} \nonumber
\hat{\mathbf{E}}'\left( \mathbf{r},\omega ,t \right) = \text{e}^{\text{i}\omega t}\hat{\mathbf{E}}\left( \mathbf{r},\omega ,t \right) \\
\hat A'_{mn}\left( t \right) = f_{mn}^A\left(  - t \right)\hat A_{mn}\left( t \right)
\end{align}
where:
\begin{equation}
f_{mn}^A\left( t \right) = \mathrm{e}^{\big[ \mathrm{i} \omega _{mn}^A - \left(
\Gamma _n^A + \Gamma _m^A \right)/2 \big]t}.
\label{f}
\end{equation}
The new dynamical variables change on the time-scale of the interaction and have the following commutator with the total Hamiltonian (See Eq. \ref{eqn}, \ref{eqn21}):
\begin{align}\nonumber
\left[ \hat{\mathbf{E}}_{\slashed{A}}'\left( \mathbf{r},\omega ,t \right),\hat{H}\left( t \right) \right] =  - \frac{\hbar \mu _0}{\pi }\text{e}^{\text{i}\omega t}&\\\nonumber
 \times \sum\limits_{m,n} f_{mn}^B\left( t \right)\omega ^2 \text{Im}\tens{G}\left( \mathbf{r},\mathbf{r}_B,\omega  \right) \mathbf{d}_{mn}^B\hat B_{mn}'\left( t \right),\\\nonumber
\left[ \hat A_{mn}'\left( t \right),\hat H\left( t \right) \right] =-  \int\limits_0^\infty  \text{d}\omega  f_{mn}^A\left(  - t \right)&\\\nonumber
  \times \left(\text{e}^{ - \text{i}\omega t} \mathbf{\hat K}_{mn}^A\left( t \right) \cdot \mathbf{\hat E}{'}_\slashed{A}\left( \mathbf{r}_A,\omega ,t \right) \right.\\
  \left.  + \text{e}^{\text{i}\omega t} \mathbf{\hat E}{'}_\slashed{A}^\dag \left( \mathbf{r}_A,\omega ,t \right) \cdot \mathbf{\hat K}_{mn}^A\left( t \right) \right)
  \label{comm}
\end{align}
where:
\begin{multline}
\mathbf{\hat K}_{mn}^A\left( t \right) =\sum\limits_k \left( \hat A'_{mk}\left( t \right)f_{mk}^A\left( t \right)\mathbf{d}_{nk}^A \right. \\
\left.  - \hat A'_{kn}\left( t \right)f_{kn}^A\left( t \right)\mathbf{d}_{km}^A \right)
\end{multline}
and a normal ordering prescription is used.

From Eqs.~(\ref{comm}) we see that the commutator between the
Hamiltonian and a dynamical variable increases the number
of electric dipole moments by one.
Hence Eq.~(\ref{eqrho}) represents a perturbative expansion of the
force with the dipole element as perturbative parameter.
In particular the electric vdW $N$-body force $\mathbf{F}_A\left( \mathbf{r}_A,\mathbf{r}_1....,\mathbf{r}_{N - 1} \right)$ acting on A due to the other atoms, with positions $\mathbf{r}_1....,\mathbf{r}_{N - 1}$, contains $2N$ electric dipole matrix elements; this force results from the application of $2N-1$ commutators:
\begin{multline}
\mathbf{F}_A\left( \mathbf{r}_A,\mathbf{r}_1....,\mathbf{r}_{N - 1},t \right)\\
=\left( \frac{1}{\text{i}\hbar } \right)^{2N-1} \int\limits_{t_0}^t \text{d} t_1\int\limits_{t_0}^{t_1} \text{d} t_2...\int\limits_{t_0}^{t_{2N-2}} \text{d} t_{2N-1}\\
\times \Big\langle I \Big| \Bigg[ \bigg[   ...\Big[ \big[
\mathbf{\hat F}_A\left( t \right) \big|_{t \to t_1},\hat H\left( t_1
\right) \big]_{t_1 \to t_2}, \hat H\left( t_2 \right) \Big]_{t_2 \to
t_3},\\
...,\hat H\left( t_{2N - 2} \right) \bigg]_{t_{2N - 2} \to t_{2N - 1}},\hat H\left( t_{2N - 1} \right)  \Bigg]_{t_{2N - 1} \to t_0}\Big| I \Big\rangle 
\end{multline}
where the expectation value is taken over the atomic+\- field free state $\big| I \big\rangle $.
This  approximate solution to the coupled dynamics is equivalent to
an iterative use of the atom--field equations, and it is valid for
weak coupling between atoms and field.

 The expectation value on free atomic and field states can be easily
performed, since after the limit $t_{2N-1} \to t_0$ the resulting
operators are evaluated at the same initial time $t_0$, which
represents the time at which the electric field and the atoms are
uncoupled.

\subsection{Van der Waals interaction between two atoms}
We consider now the vdW interaction between two atoms.
We suppose that the atomic states are incoherent
superpositions of energy eigenstates $\left| n^A \right\rangle $ and
$\left| l^B \right\rangle $ and the state of the field
is the ground state.

In normal ordering, the force operator (see Eq. \ref{multham}) can be
expressed in terms of the new dynamical variables:
\begin{multline}
\mathbf{\hat F}_A\left( t \right) =   \sum\limits_{m,n} \int\limits_0^\infty  \text{d} \omega f_{mn}^A\left( t \right)\text{e}^{ - \text{i}\omega t}\\
 \times \nabla  \big\{ \hat A'_{mn}\left( t \right)\mathbf{d}_{mn}^A \cdot \mathbf{\hat E'}_{\slashed{A}}\left( \mathbf{r},\omega ,t \right) \big\}_{\mathbf{r} = \mathbf{r}_A} +\text{h.c.}
\end{multline}
The two-body vdW interaction, which contains four electric dipole moments, involves three commutators:
\begin{multline}
\textbf{F}_A\left( \mathbf{r}_A,\mathbf{r}_B,t \right) = \left( \frac{1}{\text{i}\hbar } \right)^3 \int\limits_{t_0}^t \text{d} t_1\int\limits_{t_0}^{t_1} \text{d} t_2\int\limits_{t_0}^{t_2} \text{d} t_3\\
 \times \bigg\langle \bigg[ \Big[ \big[  \hat{\mathbf{F}}_A\left( t \right) \big|_{t \to t_1}, \hat H \left( t_1 \right) \big]_{t_1 \to t_2}\\
 ,\hat H\left( t_2 \right) \Big]_{t_2 \to t_3}, \hat H\left( t_3 \right) \bigg]_{t_3 \to t_0} \bigg\rangle 
\end{multline}

With the help of Eqs. (\ref{comm}), the commutators can be evaluated.
For example the application of one and two commutators gives:
\begin{widetext}
\begin{align}\nonumber
\frac{1}{\text{i}\hbar}\big[ \hat{\mathbf{F}}_A&\left( t \right)\big|_{t \to t_1},\hat H\left( t_1 \right) \big] =\frac{\text{i}}{\hbar} \sum\limits_{m,n} \int\limits_0^\infty  \text{d}  \omega \int\limits_0^\infty  \text{d}\omega' \text{e}^{ - \text{i}\omega t} f_{mn}^A\left( t - t_1 \right)\nabla \Big\{ \big( \text{e}^{ - \text{i}\omega' t_1} \mathbf{\hat K}_{mn}^A\left( t_1 \right) \cdot \hat{\mathbf{E}}'_\slashed{A}\left( \mathbf{r}_A,\omega' ,t_1 \right) + \\\nonumber
 & + \text{e}^{\text{i}\omega' t_1} \mathbf{\hat E}{'}_\slashed{A}^\dag  \left( \mathbf{r}_A,\omega' ,t_1 \right) \cdot  \mathbf{\hat K}_{mn}^A\left( t_1 \right) \big) \mathbf{d}_{mn}^A \cdot \hat{\mathbf{E}}'_{\slashed{A}}\left( \mathbf{r},\omega ,t_1 \right) \Big\} _{\mathbf{r} = \mathbf{r}_A}\\\nonumber
&+ \frac{\text{i} \mu _0}{\pi }\sum\limits_{m,n,r,s} \int\limits_0^\infty  \text{d}\omega \omega^2 \text{e}^{ - \text{i}\omega \left( t - t_1 \right)}f_{mn}^A\left( t \right)f_{rs}^B\left( t_1 \right)\nabla \Big\{ \hat A'_{mn}\left( t_1 \right)  \hat B'_{rs}\left( t_1 \right) \mathbf{d}_{mn}^A \cdot \operatorname{Im} \tens{G}\left( \mathbf{r},\mathbf{r}_B,\omega  \right) \cdot \mathbf{d}_{rs}^B \Big\}_{\mathbf{r} = \mathbf{r}_A}+ \text{h.c.}\\\nonumber
\left( \frac{1}{\text{i}\hbar } \right)^2\Big[ \big[  \mathbf{\hat F}_A&\left( t \right) \big|_{t \to t_1},\hat H\left( t_1 \right) \big]_{t_1 \to t_2},\hat H\left( t_2 \right) \Big] =-\frac{\mu _0}{\pi \hbar}\sum\limits_{m,n,r,s} \int\limits_0^\infty  \text{d}\omega  \int\limits_0^\infty  \text{d}\omega' f_{mn}^A\left( t - t_1 \right)f_{rs}^B\left( t_2 \right)\\\nonumber
& \times \nabla \Big\{ \omega {'^2}\text{e}^{ - \text{i}\omega t}\Big[ \text{e}^{ - \text{i}\omega '\left( t_1 - t_2 \right)}\mathbf{\hat K}_{mn}^A\left( t_1 \right) \big|_{t_1 \to t_2} \cdot \operatorname{Im} \tens{G}\left( \mathbf{r}_A,\mathbf{r}_B,\omega ' \right) \cdot \mathbf{d}_{rs}^B \hat B'_{rs}\left( t_2 \right) \\\nonumber
& - \text{e}^{\text{i}\omega '\left( t_1 - t_2 \right)}\hat B'_{rs}\left( t_2 \right)\mathbf{d}_{rs}^B \cdot \operatorname{Im} \tens{G}\left( \mathbf{r}_B,\mathbf{r}_A,\omega ' \right) \cdot \mathbf{\hat K}_{mn}^A\left( t_1 \right) \big|_{t_1 \to t_2} \Big]\mathbf{d}_{mn}^A \cdot \hat{\mathbf{E}}'_\slashed{A}\left( \mathbf{r},\omega,t_2 \right)\\\nonumber
 & + \omega ^2 \text{e}^{ - \text{i}\omega \left( t - t_2 \right)}\Big[ \text{e}^{ - \text{i}\omega 't_1}\mathbf{\hat K}_{mn}^A\left( t_1 \right) \big|_{t_1 \to t_2} \cdot \hat{\mathbf{E}}'_\slashed{A} \left( \mathbf{r}_A,\omega ',t_2 \right) + \\\nonumber
&+  \text{e}^{\text{i}\omega 't_1}\mathbf{\hat E}{'}_\slashed{A}^\dag \left( \mathbf{r}_A,\omega ',t_2 \right) \cdot \mathbf{\hat K}_{mn}^A\left( t_1 \right) \big|_{t_1 \to t_2} \Big]  \mathbf{d}_{mn}^A \cdot \operatorname{Im} \tens{G}\left( \mathbf{r},\mathbf{r}_B,\omega  \right) \cdot \mathbf{d}_{rs}^B \hat B'_{rs}\left( t_2 \right) \Big\}_{\mathbf{r} = \mathbf{r}_A}\\\nonumber
 & -\frac{\mu _0}{\pi \hbar}\sum\limits_{m,n,r,s} \int\limits_0^\infty  \text{d}\omega  \int\limits_0^\infty  \text{d}\omega ' \text{e}^{ - \text{i}\omega \left( t - t_1 \right)}\omega ^2 \nabla \Big\{ \big[ f_{mn}^A\left( t - t_2 \right)f_{rs}^B\left( t_1 \right)\big( \text{e}^{ - \text{i}\omega 't_2} \mathbf{\hat K}_{mn}^A\left( t_2 \right) \cdot \hat{\mathbf{E}}'_\slashed{A}\left( \mathbf{r}_A,\omega ',t_2 \right)\\\nonumber
&  + \text{e}^{\text{i}\omega 't_2} \mathbf{\hat E}{'}_\slashed{A}^\dag \left( \mathbf{r}_A,\omega ',t_2 \right) \cdot  \mathbf{\hat K}_{mn}^A\left( t_2 \right) \big) \hat B'_{rs}\left( t_2 \right)+ f_{mn}^A\left( t \right)f_{rs}^B\left( t_1 - t_2 \right) \hat A'_{mn} \left( t_2 \right)\\\nonumber
&\times \big( \text{e}^{ - \text{i}\omega 't_2} \mathbf{\hat K}_{rs}^B\left( t_2 \right) \cdot \hat{\mathbf{E}}'_\slashed{B}\left( \mathbf{r}_B,\omega ',t_2 \right) +  \text{e} ^{\text{i}\omega 't_2}\mathbf{\hat E}{'}_\slashed{B}^\dag \left( \mathbf{r}_B,\omega ',t_2 \right) \cdot \mathbf{\hat K}_{rs}^B\left( t_2 \right) \big) \Big]\\
&\times  \mathbf{d}_{mn}^A \cdot \operatorname{Im} \tens{G}\left( \mathbf{r},\mathbf{r}_B,\omega  \right) \cdot \mathbf{d}_{rs}^B \Big\}_{\mathbf{r} = \mathbf{r}_A}+\text{h.c.}
\end{align}
\end{widetext}
The commutators between $\mathbf{\hat K}_{mn}$ and the Hamiltonian have not been considered since they lead to higher order corrections in the
 electric dipole $\textbf{d}^A$ and $\textbf{d}^B$. 
 
  We then evaluate the last commutator and take the expectation value on the atomic and field states. The thermal expectation value over the free field variables can be performed with the help of the following fluctuation relations for zero-temperature \cite{Buhmann,Buhmann2}:
\begin{align}\nonumber
\Big\langle \mathbf{\hat E}^{\left( 0 \right)}\left( \mathbf{r},\omega
,t \right)
\mathbf{\hat E}^{\left( 0 \right)\dag }\left( \mathbf{r}',\omega ',t \right) \Big\rangle&  \\ 
&\hspace{-4cm} = \frac{\hbar \mu _0}{\pi }
 \textup{Im}\tens{G}\left( \mathbf{r},\mathbf{r}',\omega  \right)
 \omega ^2\delta \left( \omega  - \omega ' \right)
\end{align}
  After some algebra we obtain:
\begin{widetext}
\begin{multline}
\mathbf{F}\left( \mathbf{r}_A,\mathbf{r}_B,t \right) =  - \frac{i\mu _0^2}{2\pi ^2\hbar }\sum\limits_{n,l} p_n^A(t) p_l^B(t) \sum\limits_{k,p} \int\limits_0^\infty  \text{d}\omega  \int\limits_0^\infty  \text{d}\omega '\omega ^2\omega {'^2}\nabla _A\mathbf{d}_{nk}^A \cdot \text{Im}\tens{G}\left( \mathbf{r}_A,\mathbf{r}_B,\omega  \right) \cdot \mathbf{d}_{pl}^B\\
 \times \mathbf{d}_{lp}^B \cdot \text{Im}\tens{G}\left( \mathbf{r}_B,\mathbf{r}_A,\omega ' \right) \cdot \mathbf{d}_{kn}^A\int\limits_{t_0}^t \text{d} t_1\int\limits_{t_0}^{t_1} \text{d} t_2\int\limits_{t_0}^{t_2} \text{d} t_3\\
\times \left\{ \text{e}^{ - \text{i}\omega \left( t - t_1 \right)} \left( f_{nk}^A\left( t - t_2 \right) - f_{kn}^A\left( t - t_2 \right) \right)\left( \text{e}^{ - \text{i}\omega '\left( t_2 - t_3 \right)}f_{lp}^B\left( t_1 - t_3 \right) - \text{e}^{\text{i}\omega '\left( t_2 - t_3 \right)}f_{pl}^B\left( t_1 - t_3 \right) \right) \right.\\
+ \text{e}^{ - \text{i}\omega \left( t - t_1 \right)}\left( \text{e}^{ - \text{i}\omega '\left( t_2 - t_3 \right)}f_{nk}^A\left( t - t_3 \right) - \text{e}^{\text{i}\omega '\left( t_2 - t_3 \right)}f_{kn}^A\left( t - t_3 \right) \right)\left( f_{lp}^B\left( t_1 - t_2 \right) - f_{pl}^B\left( t_1 - t_2 \right) \right)\\
 + \text{e}^{ - \text{i}\omega \left( t - t_2 \right)}\left( f_{nk}^A\left( t - t_1 \right) - f_{kn}^A\left( t - t_1 \right) \right)\left( \text{e}^{ - \text{i}\omega '\left( t_1 - t_3 \right)}f_{lp}^B\left( t_2 - t_3 \right) - \text{e}^{\text{i}\omega '\left( t_1 - t_3 \right)}f_{pl}^B\left( t_2 - t_3 \right) \right)\\
  +\left. \text{e}^{ - \text{i}\omega \left( t - t_3 \right)}\left( \text{e}^{ - \text{i}\omega '\left( t_1 - t_2 \right)} - \text{e}^{\text{i}\omega '\left( t_1 - t_2 \right)} \right)\left( f_{nk}^A\left( t - t_1 \right) - f_{kn}^A\left( t - t_1 \right) \right)f_{lp}^B\left( t_2 - t_3 \right ) \right\} +\text{c.c.} 
  \label{formula}
\end{multline}
\end{widetext}
where $\nabla_A$ is now applied to both Green's tensors (after
exploiting their symmetry and introducing a factor $1/2$). The
function $f$ was defined in Eq. (\ref{f}) and
$p_n^A(t) = \big\langle \hat A_{nn}\left( t \right) \big\rangle $
and $p_l^B(t) = \big\langle \hat B_{ll}\left( t \right) \big\rangle 
$ represent the atomic populations of
the states $\left| n \right\rangle $ and $\left| l
\right\rangle $. We have considered time-reversal symmetric systems where $\textbf{d}_{mn}$ is real ($\textbf{d}_{mn}=\textbf{d}_{nm}$), and reciprocal media  ($\tens{G}^{\text{T}}\left( \mathbf{r}_A,\mathbf{r}_B,\omega  \right)=\tens{G}\left( \mathbf{r}_B,\mathbf{r}_A,\omega  \right)$).

With the exception of resonant cavity-QED scenarios, we can assume
the quantity 
$\omega^2 \textup{Im}\tens{G}\left(
\mathbf{r}_B,\mathbf{r}_A,\omega  \right)$ to be
sufficiently flat and to not exhibit any narrow peaks in vicinity
of any atomic frequency (weak coupling). 
For weak coupling, we may evaluate the time-integral by means of the Markov approximation, extending the lower limit of the time integral to $t_0=-\infty$.
The resulting integrals are not converging. In order to force the convergence we add an infinitesimal factor to the frequency $\omega$, 
$\omega \to \omega - \text{i} \epsilon$, where $\epsilon>0$. Note that the opposite sign convention for this infinitesimal factor would lead to divergent integrals.
Time-integration leads to the energy denominators in Table 1 in the main text.

The frequency denominators can be combined:
\begin{align}\nonumber
 1/D_2 + 1/D_7 + 1/D_{10} \\ \nonumber 
=1/( \omega ^{(  - )} - \omega ' \big)( \omega ' + \omega _{kn}^{A(  - )} )( \omega ' + \omega _{pl}^{B(  + )} )\\\nonumber
1/D_3 + 1/D_6 + 1/D_{11} \\ \nonumber
=1/(\omega ^{( - )} + \omega ')(\omega ' + \omega _{kn}^{A( + )})(\omega ' + \omega _{pl}^{B( - )}) \\ \nonumber
1/D_1 + 1/D_9= \frac{1}{(\omega ^{(  - )} + \omega ')(\omega _{kn}^{A(  -  )} + \omega _{pl}^{B( -  )})}\\ \nonumber
\times \bigg( \frac{1}{\omega ^{(  -  )} + \omega _{kn}^{A( -  )}} + \frac{1}{\omega ' + \omega _{pl}^{B(  -  )}}\bigg) \\\nonumber
1/D_4 + 1/D_{12}= \frac{1}{(\omega ' -\omega ^{( - )})(\omega _{kn}^{A( + )} + \omega _{pl}^{B( + )})} \\ \nonumber 
 \times \bigg( \frac{1}{\omega ^{( - )} - \omega _{kn}^{A( + )}} - \frac{1}{\omega ' + \omega _{pl}^{B( + )}} \bigg)\\\nonumber
1/D_5 = \frac{1}{(\omega ^{( - )} - \omega ')(\omega _{kn}^{A( - )} + \omega _{pl}^{B( - )})}\\\nonumber
 \times \bigg( \frac{1}{\omega ' + \omega _{kn}^{A( - )}} - \frac{1}{\omega ^{\left(  -  \right)} + \omega _{kn}^{A( - )}} \bigg)\\\nonumber
1/D_8=\frac{1}{(\omega ^{( - )} + \omega ')(\omega _{kn}^{A( + )} + \omega _{pl}^{B( + )})}\\ 
 \times \bigg( \frac{1}{\omega ' + \omega _{kn}^{A( + )}} + \frac{1}{\omega ^{\left(  -  \right)} - \omega _{kn}^{A( + )}} \bigg),
\end{align}
which implies:
\begin{multline}
\sum\limits_{i = 1}^{16} {\frac{1}{{{D_i}}}}  + {\text{c.c.}} = f_1\left( {\omega '} \right)\left( {\frac{1}{{{\omega ^{\left(  -  \right)}} + \omega '}} + \frac{1}{{{\omega ^{\left(  + \right)}} - \omega '}}} \right)\\
 + f_2\left( \omega^{(-)}  \right)\left( {\frac{1}{{\omega ' + {\omega ^{\left(  -  \right)}}}} + \frac{1}{{\omega ' - {\omega ^{\left(  -  \right)}}}}} \right)+ {\text{c.c.}} 
 \label{eq50}
\end{multline}
where we have defined the following functions:
\begin{align}\nonumber
f_1\left( \xi  \right) =& \frac{1}{\left( \omega _{kn}^{A\left(  +  \right)} + \omega _{pl}^{B\left(  +  \right)} \right)\left( \xi  + \omega _{kn}^{A\left(  +  \right)} \right)}\\\nonumber
& + \frac{1}{\left( \omega _{kn}^{A\left(  -  \right)} + \omega _{pl}^{B\left(  -  \right)} \right)\left( \xi  + \omega _{pl}^{B\left(  -  \right)} \right)} \\\nonumber
&+ \frac{1}{\left( \xi  + \omega _{kn}^{A\left(  +  \right)} \right)\left( \xi  + \omega _{pl}^{B\left(  -  \right)} \right)} \\\nonumber
f_2\left( \xi  \right) =& \frac{1}{\left( \omega _{kn}^{A\left(  +  \right)} + \omega _{pl}^{B\left(  +  \right)} \right)\left( \xi  - \omega _{kn}^{A\left(  +  \right)} \right)}\\\nonumber
& + \frac{1}{\left( \omega _{kn}^{A\left(  -  \right)} + \omega _{pl}^{B\left(  -  \right)} \right)\left( \xi  + \omega _{kn}^{A\left(  -  \right)} \right)}\\
&+ \left( \frac{1}{\xi  + \omega _{kn}^{A\left(  -  \right)}} - \frac{1}{\xi  - \omega _{kn}^{A\left(  +  \right)}} \right)\frac{1}{\xi  + \omega _{pl}^{B\left(  -  \right)}}
 \end{align}
 and $\omega _{kn}^{A\left(  \pm  \right)}
=  \omega_{kn}^A \pm \text{i}\left( \Gamma _k^A + \Gamma _n^A \right)/2$ , $\omega _{pl}^{B\left(  \pm  \right)}
= \omega _{pl}^B \pm \text{i}\left( \Gamma _p^B + \Gamma _l^B \right)/2$, $\omega ^{\left(  \pm  \right)} = \omega  \pm \text{i}\epsilon$.

For the first term in Eq. (\ref{eq50}) we integrate over $\omega$ and for the second term we integrate over $\omega'$. We use the
 identity $\textup{Im}\tens{G} = \left( \tens{G} - \tens{G}^* \right)/2\mathrm{i}$
 and the Schwarz reflection principle for the Green tensor:
\begin{gather} \nonumber 
\int\limits_0^{  \infty } \mathrm{d}\omega '\omega '^2 \left( \frac{1}{\omega'  +
 \omega ^{(-)} } + \frac{1}{\omega ' - \omega^{(-)} } \right)
\textup{Im}\tens{G}\left( \mathbf{r}_A,\mathbf{r}_B,\omega ' \right)\\
=\frac{1}{2\mathrm{i}}\int\limits_{ - \infty }^{  \infty } \mathrm{d}\omega '\omega '^2
 \left( \frac{1}{\omega'  + \omega^{(-)} } + \frac{1}{\omega ' - \omega^{(-)} } \right)
 \tens{G}\left( \mathbf{r}_A,\mathbf{r}_B,\omega ' \right)
\end{gather}
The Green's tensor is analytic in the upper half of the complex plane,
including the real axis, and it is also finite at the origin.
We close the path with an infinitely large half-circle  in the upper complex half-plane and take the residuum inside the path. The
integral along the infinite  semi-circle vanishes for $\textbf{r}_A
\ne\textbf{r}_B$ because:
\begin{equation}\mathop {\lim }\limits_{\left| \omega  \right| \to  +
\infty }
\omega ^2\left. \tens{G}\left( \mathbf{r}_A,\mathbf{r}_B,\omega  \right)
\right|_{\mathbf{r}_A \ne \mathbf{r}_B} = 0\end{equation}
We thus find:
\begin{multline}\int\limits_0^{  \infty } \mathrm{d}\omega '\omega '^2 \left(
 \frac{1}{\omega'  + \omega^{(-)} } + \frac{1}{\omega ' - \omega^{(-)} } \right)
 \textup{Im}\tens{G}\left( \mathbf{r}_A,\mathbf{r}_B,\omega ' \right) \\
 \simeq \pi \omega ^2 \tens{G}\left( \mathbf{r}_A,\mathbf{r}_B,- \omega  \right)
  \label{int}\end{multline}
  The total force can be expressed as sum of two terms:
\begin{equation}
\mathbf{F}_A\left( \mathbf{r}_A,\mathbf{r}_B,t \right) = \mathbf{F}_A^1\left( \mathbf{r}_A,\mathbf{r}_B,t \right) + \mathbf{F}_A^2\left( \mathbf{r}_A,\mathbf{r}_B,t \right)\nonumber
\end{equation}
\begin{align}\nonumber
\mathbf{F}_A^1\left( \mathbf{r}_A,\mathbf{r}_B,t \right) =& \frac{\mu _0^2}{2\pi \hbar }\sum\limits_{n,l} p_n^A(t) p_l^B(t) \sum\limits_{k,p} \int\limits_0^\infty  \text{d} \omega \omega ^4\\ \nonumber
&\hspace{-1.5cm} \times \nabla _A\text{Im}\left\{ \left( \mathbf{d}_{nk}^A \cdot \tens{G}\left( \mathbf{r}_A,\mathbf{r}_B,\omega  \right)\cdot \mathbf{d}_{pl}^B \right)^2g_1\left( \omega  \right) \right\}\\\nonumber
\mathbf{F}_A^2\left( \mathbf{r}_A,\mathbf{r}_B,t \right) =&\frac{\mu _0^2}{2\pi  \hbar }\sum\limits_{n,l} p_n^A(t) p_l^B(t) \sum\limits_{k,p} \int\limits_0^\infty  \text{d} \omega \omega ^4\\
&\hspace{-1.5cm} \times \nabla _A\left\{ \left| \mathbf{d}_{nk}^A \cdot \tens{G}\left( \mathbf{r}_A,\mathbf{r}_B,\omega  \right) \cdot \mathbf{d}_{pl}^B \right|^2 g_2 \left( \omega  \right) \right\}
\label{eqn55}
\end{align}
  where:
  \begin{multline}
g_1\left( \omega  \right) = f_1^*\left( \omega  \right) + f_2^*\left( \omega  \right)=\\
\frac{1}{\omega  + \omega _{pl}^{B\left(  +  \right)}}\left( \frac{1}{\omega  + \omega _{kn}^{A\left(  +  \right)}} \right.\\
\left.  + \frac{1}{\omega  + \omega _{kn}^{A\left(  -  \right)}} - \frac{1}{\omega  - \omega _{kn}^{A\left(  -  \right)}} \right)\\
 + \frac{1}{\omega _{kn}^{A\left(  +  \right)} + \omega _{pl}^{B\left(  +  \right)}}\left( \frac{1}{\omega  + \omega _{kn}^{A\left(  +  \right)}} + \frac{1}{\omega  + \omega _{pl}^{B\left(  +  \right)}} \right)\\
  + \frac{1}{\omega _{kn}^{A\left(  -  \right)} + \omega _{pl}^{B\left(  -  \right)}}\left( \frac{1}{\omega  + \omega _{kn}^{A\left(  -  \right)}} + \frac{1}{\omega  - \omega _{kn}^{A\left(  -  \right)}} \right)
\end{multline}
  and:
    \begin{multline}
g_2\left( \omega  \right) = \text{Im} \left[ f_1\left( \omega  \right) + f_2\left( \omega  \right) \right]\\
 =2 \operatorname{Re} \left[ \frac{1}{\omega _{kn}^{A\left(  +  \right)} + \omega } + \frac{1}{\omega _{kn}^{A\left(  +  \right)} - \omega } \right]\text{Im}\frac{1}{\omega  + \omega _{pl}^{B\left(  -  \right)}}
\end{multline}
We consider then the limiting case of vanishing line-widths:
\begin{align}\nonumber
\varepsilon _A =& \left( \Gamma _n^A + \Gamma _k^A \right)/2 \to 0^ + \\
\varepsilon _B =& \left( \Gamma _l^A + \Gamma _p^B \right)/2 \to 0^ + 
\end{align}
In this limit the function $g_1$  can be simplified:
\begin{multline}\mathop {\lim }\limits_{\epsilon_{A,B} \to 0^ + } g_1\left( \omega  \right) = \frac{4\left( \omega  - \omega _{kn}^A \right)\left( \omega  + \omega _{kn}^A \right)}{\left( \left( \omega  + \omega _{kn}^A \right)^2 + \varepsilon _A^2 \right)\left( \omega  - \omega _{kn}^A + \text{i}\varepsilon _A \right)}\\
 \times \frac{\left( \omega  + \omega _{kn}^A + \omega _{pl}^B \right)}{\left( \omega  + \omega _{pl}^B + \text{i}\varepsilon _B \right)\left( \omega _{kn}^A + \omega _{pl}^B \right)}\end{multline}
Using the property $\frac{1}{x \pm \text{i}\varepsilon } = \mathcal{P}\frac{1}{x} \mp \text{i}\pi \delta \left( x \right)$, where $\mathcal{P}$ is the principal value, we can also simplify $g_2$ :
 \begin{equation}
 g_2\left( \omega  \right) =2 \pi \operatorname{Re} \Bigg[ \frac{1}{\omega _{kn}^{A\left(  +  \right)} + \omega } + \frac{1}{\omega _{kn}^{A\left(  +  \right)} - \omega } \Bigg] \delta \left( {\omega  - \omega _{lp}^B} \right)
 \end{equation}

With these results, after performing a Wick rotation on the imaginary
axis we find the following non resonant and resonant contributions to
the $\textbf{F}_A^1$:
\begin{multline}
\mathbf{F}_A^1\left( \mathbf{r}_A,\mathbf{r}_B,t \right) = \frac{\mu
_0^2}{2\pi \hbar }\sum\limits_{n,l} p_n^A(t) p_l^B(t)
\sum\limits_{k,p} \int\limits_0^\infty  \text{d} \xi  \xi ^4\\
\times \frac{g_1\left( \text{i}\xi  \right) + g_1^*\left(  -
\text{i}\xi  \right)}{2}
\nabla _A\left\{ \left( \mathbf{d}_{nk}^A \cdot \tens{G}\left(
\mathbf{r}_A,\mathbf{r}_B,\text{i}\xi  \right) \cdot \mathbf{d}_{pl}^B
\right)^2 \right\}\\
+ \frac{\mu _0^2}{2\hbar }\sum\limits_{n,l} p_n^A(t) p_l^B(t)
\sum\limits_{k,p}\\
\times \nabla_A \Big\{ \text{Res}_1 \big[ g_1\left( \omega
\right)\omega ^4 \left( \mathbf{d}_{nk}^A \cdot \tens{G}\left(
\mathbf{r}_A,\mathbf{r}_B,\omega  \right) \cdot \mathbf{d}_{pl}^B
\right)^2 \big]  \\
 - \text{Res}_2 \big[ g_1^*\left(  - \omega  \right)\omega ^4 \left(
\mathbf{d}_{nk}^A \cdot \tens{G}\left(
\mathbf{r}_A,\mathbf{r}_B,\omega  \right) \cdot \mathbf{d}_{pl}^B
\right)^2 \big] \Big\}
\end{multline}
where $\text{Res}_1$ indicates the sum of the residues in the first
quadrant, and $\text{Res}_2$ the sum of the residues in the second
quadrant. Similarly, $\textbf{F}_A^2$ reduces to
 \begin{multline}
 \mathbf{F}_A^2\left( \mathbf{r}_A,\mathbf{r}_B,t \right) = \mu _0^2
\sum\limits_{l } p_l^B(t)\sum\limits_{p < l} \left( \omega _{lp}^B
\right)^4 \nabla _A\\
  \times \left\{ \mathbf{d}_{lp}^B \cdot \tens{G}\left(
\mathbf{r}_B,\mathbf{r}_A,\omega _{lp}^B \right) \cdot \bm{\alpha}_A\left(
\omega _{lp}^B \right) \cdot {\tens{G}^*}\left(
\mathbf{r}_A,\mathbf{r}_B,\omega _{lp}^B \right)\cdot
\mathbf{d}_{pl}^B \right\}\\
 \end{multline}
 where $\bm{\alpha}_A$ is the polarizability of the excited atom A.
The sum of $\textbf{F}_A^1$ and $\textbf{F}_A^2$ gives the non-resonant
and resonant contributions of the total vdW force, see Eqs.
(\ref{nonresonant}) and (\ref{resonant}).


\newpage 
\begin{thebibliography}{10}

\bibitem{Casimir48}
H. B. G. Casimir, \textit{Proc. K. Ned. Akad. Wet.}
\textbf{51}, 793 (1948).

\bibitem{CasimirPolder48}
H. B. G. Casimir and D. Polder, \textit{Phys. Rev.}
\textbf{73}, 360 (1948).

\bibitem{kihara}
T. Kihara, \textit{Intermolecular forces} (John Wiley \& Sons, New
York, 1977).

\bibitem{nir}
S. Nir, \textit{Prog. Surf. Sci.} \textbf{8}, 1 (1976).

\bibitem{pater}
J. de Pater, J.J. Lissauer
\textit{Planetary Sciences}, Cambridge University Press, Cambridge
(2010).

\bibitem{Preto}
J. Preto, M. Pettini and J. A. Tuszynski, \textit{Phys. Rev. E}
\textbf{91}, 052710 (2015).

\bibitem{Serry}
F. M. Serry, D. Walliser and G. J. Maclay, \textit{J. Appl. Phys.}
\textbf{84}, 2501 (1998).

\bibitem{wilson}
M. A. Wilson, P. Bushev, J. Eschner, F. Schmidt-Kaler, C. Becher, R.
Blatt, and U. Dorner, \textit{Phys. Rev. Lett.}
 \textbf{91}, 213602 (2003).
 
\bibitem{bushev}
P. Bushev, A. Wilson, J. Eschner, C. Raab, F. Schmidt-Kaler, C.
Becher, and R. Blatt, \textit{Phys. Rev. Lett.}
\textbf{92}, 223602 (2004).

\bibitem{Buhmann}
S. Y. Buhmann, \emph{Dispersion forces I} (Springer, Heidelberg,
2012).

\bibitem{power3}
E. A. Power, T. Thirunamachandran, \textit{Phys. Rev. A}
 \textbf{47}, 2539 (1993).
 
\bibitem{Gomberoff}
L. Gomberoff, R. R. McLone, E.  A. Power, \textit{J. Chem. Phys.}
 \textbf{44}, 4148 (1966).
 
\bibitem{McLone}
R. R. McLone, E. A. Power, \textit{Proc. R. Soc. Lond. Ser. A}
 \textbf{286}, 573 (1965).
 
\bibitem{Philpott}
M. R. Philpott, \textit{Proc. Phys. Soc. Lond.}
 \textbf{87}, 619 (1966).
 
 \bibitem{power}
E. A. Power, T. Thirunamachandran, \textit{Phys. Rev. A}
 \textbf{51}, 3660 (1995).
 
\bibitem{power2}
E. A. Power, T. Thirunamachandran, \textit{Chem. Phys.}
 \textbf{171}, 1 (1993).
 
 \bibitem{sherk}
Y. Sherkunov, \textit{Phys. Rev. A}
 \textbf{75}, 012705 (2007).
 
 \bibitem{manuel}
M. Donaire, R. Gu\'{e}rout, A. Lambrecht, \textit{Phys. Rev. Lett.}
\textbf{115}, 033201 (2015).

 \bibitem{berman}
P. R. Berman, \textit{Phys. Rev. A}
 \textbf{91}, 042127 (2015).
 
 \bibitem{rizzuto}
L. Rizzuto, R. Passante, F. Persico, \textit{Phys. Rev. A}
 \textbf{70}, 012107 (2004).
 
\bibitem{Haakh}
H.R. Haakh, J. Schiefele, and C. Henkel, \textit{Int. J. Mod. Phys.:
Conf. Ser.} \textbf{14}, 347 (2012).

\bibitem{milonni3}
P.W. Milonni and S. M. H. Rafsanjani, \textit{Phys. Rev. A } \textbf{92}, 062711 (2015).

\bibitem{manuel2}
M. Donaire, \textit{ arXiv:1603.08195 } (2016).

\bibitem{manuel3}
M. Donaire, \textit{ arXiv:1604.07071 } (2016).

\bibitem{Scheel15}
S. Scheel, S. Y. Buhmann, C. Clausen, and P. Schnee\-weiss
Phys. Rev. A \textbf{92}, 043819 (2015).

\bibitem{safari}
H. Safari and M. R. Karimpour, \textit{Phys. Rev. Lett.}
 \textbf{114}, 013201 (2015).
 

\bibitem{Buhmann-Welsch}
S. Y. Buhmann and D. G. Welsch, \textit{Prog. Quantum Electron.}
\textbf{31}, 51 (2007).

\bibitem{Buhmann2}
S. Y. Buhmann, \emph{Dispersion Forces II}
(Springer, Heidelberg, 2013).

\bibitem{knoll}
L.~Kn\"oll, S.~Scheel, D.-G. Welsch, \textit{QED in Dispersing
and Absorbing Media} in J.~Pe\v{r}ina (ed.) \textit{Coherence and
Statistics of Photons and Atoms}, p.~1 (Wiley, New York, 2001).

\bibitem{ackerhalt}
J. R. Ackerhalt, P. L. Knight, and J. H. Eberly, \textit{Phys. Rev.
Lett.} \textbf{30}, 456 (1973).

\bibitem{steck}D. A. Steck, Rubidium 87D Line Data, Cesium D Line Data \url{http://steck.us/alkalidata} (2009).

\bibitem{born}S. Y. Buhmann and  Welsch,  \textit{Appl. Phys. B}
\textbf{82}, 189 (2006).

\bibitem{Shamoon13}
E. Shahmoon and G. Kuritzki, Phys. Rev. A \textbf{87}, 062105 (2013).
\bibitem{Shamoon14}
E. Shahmoon, I. Mazets and G. Kurizki, Proc. Nat. Akad. Sci.
\textbf{111}, 10485 (2013).

\bibitem{ScheelHaakh}
H. R. Haakh and S. Scheel, Phys. Rev. A \textbf{91}, 052707 (2015).
\bibitem{Farina}
R. de Melo de Souza, W. J. M. Kort-Kamp, F. S. S. Rosa and C. Farina,
Phys. Rev. A \textbf{91}, 052708 (2015).

\end{thebibliography}
\end{document}